\documentclass{aa}
\usepackage{epsfig,amssymb,txfonts}

\newcommand{\dof}{\mathrm{d.o.f.}}

\begin{document}

\title{Evidence for Supernova Signatures in the Spectrum of the
Late-time Bump of the Optical Afterglow of GRB\,021211%
\thanks{Based on observations made with ESO telescopes under programmes
70.D-0412 and 270.D-5022.}}
\titlerunning{Supernova in GRB\,021211}

\author{M.~Della~Valle\inst{1} \and D.~Malesani\inst{2} \and S.~Benetti\inst{3} \and
V.~Testa\inst{4} \and M.~Hamuy\inst{5,1} \and
L.A.~Antonelli\inst{4} \and G.~Chincarini\inst{6,7} \and G.~Cocozza\inst{4,8} \and S.~Covino\inst{6}
\and P.~D'Avanzo\inst{6} \and D.~Fugazza\inst{9} \and G.~Ghisellini\inst{6} \and R.~Gilmozzi\inst{10}
\and D.~Lazzati\inst{11} \and E.~Mason\inst{10} \and P.~Mazzali\inst{12} \and L.~Stella\inst{4}
}

\authorrunning{Della~Valle et al.}
\offprints{M.~Della~Valle\\ \email{massimo@arcetri.astro.it}}

\institute{ 
INAF, Osservatorio Astrofisico di Arcetri, largo E. Fermi 5, 50125 Firenze, Italy.
\and        
International School for Advanced Studies (SISSA/ISAS), via Beirut 2-4, 34016 Trieste, Italy.
\and        
INAF, Osservatorio Astronomico di Padova, vicolo dell'Osservatorio 5, 35122 Padova, Italy.
\and        
INAF, Osservatorio Astronomico di Roma, via Frascati 33, 00040 Monteporzio Catone (Roma), Italy.
\and        
Carnegie Observatories, 813 Santa Barbara Street, Pasadena, California 91101, USA.
\and        
INAF, Osservatorio Astronomico di Brera, via E. Bianchi 46, 23807 Merate (Lc), Italy.
\and
University of Milano--Bicocca, Department of Physics, Piazza delle Scienze 3, 20126 Milano, Italy.
\and        
University of Roma ``Tor Vergata'', Department of Physics, via della Ricerca Scientifica 1, 00133 Roma, Italy.
\and        
INAF, Telescopio Nazionale Galileo, Roque de Los Muchachos, PO box 565, 38700 Santa Cruz de La Palma, Spain.
\and        
ESO, Alonso de Cordova 3107, Casilla 19001, Vitacura, Santiago, Chile.
\and        
Institute of Astrophysics, University of Cambridge, Madingley Road, CB3 0HA Cambridge, UK.
\and        
INAF, Osservatorio Astronomico di Trieste, via Tiepolo 11, 34131 Trieste, Italy.
}

\date{Received }

\abstract{We present photometric and spectroscopic observations of the
gamma-ray burst GRB\,021211 obtained during the late stages of its
afterglow. The light curve shows a rebrightening occurring $\sim
25$~days after the GRB. The analysis of a VLT spectrum obtained during
the bump (27 days after the GRB) reveals a suggestive resemblance with
the spectrum of the prototypical type-Ic SN\,1994I, obtained $\sim
10$~days past maximum light. Particularly we have measured a strong,
broad absorption feature at 3770~\AA, which we have identified with
Ca\,II blueshifted by $\sim 14\,400$~km/s, thus indicating that a
supernova (SN) component is indeed powering the `bump' in the afterglow
decay. Assuming SN\,1994I as a template, the spectroscopic and
photometric data together indicate that the SN and GRB explosions were
at most separated by a few days. Our results suggest that GRBs might be
associated also to standard type-Ic supernovae.
}

\maketitle

\section{Introduction.}\label{sec:intro}

There is growing evidence that long-duration GRBs (i.e. those lasting
more than $\sim 2$~s; e.g. Fishman et al. \cite{Fi94}) are associated
with the death of massive stars. The most compelling case for the
existence of a SN-GRB connection is represented by the spatial and
temporal coincidence between GRB\,980425 and SN\,1998bw (Galama et
al. \cite{Ga98}). Very recently, in the spectrum of the nearby ($z =
0.1687$) GRB\,030329, Stanek et al. (\cite{St03}) found supernova
features, emerging out of the afterglow spectrum and resembling those of
SN\,1998bw (Patat et al. \cite{Pa01}). Several other possible SN-GRB
associations have been suggested (e.g. Wang \& Wheeler \cite{WW98};
Woosley et al. \cite{Wo98}; Germany et al. \cite{Ge00}; Terlevich et
al. \cite{Te99}; Rigon et al. \cite{Ri03}), although none really
compelling due to the poor spatial and temporal coincidences.

The remaining evidence for the existence of a GRB-SN connection relies
upon the detection of a rebrightening in the afterglow light curves, $20
- 30$ days after the GRB (e.g. Bloom et al. \cite{Bl99},
\cite{Bl02}; Lazzati et al. \cite{La01}). These `bumps' have been
interpreted as signatures of SNe at maximum, emerging out of the
decaying afterglows. However this evidence and, consequently, the
interpretation, are based only on a few photometric measurements in the
afterglow light curves and on the assumption that the evolution of the
underlying SN is similar to SN\,1998bw, except for a luminosity
rescaling factor (see e.g. Price et al. \cite{Pr03}). Alternative
explanations, such as dust echoes (Esin \& Blandford \cite{EB00}),
thermal re-emission of the afterglow light (Waxman \& Draine
\cite{WD00}), or thermal radiation from a preexisting SN remnant (Dermer
\cite{De03}) are not yet ruled out.  Indirect evidence that at least
some long-duration GRBs are associated with the death of massive stars
is provided by the detection of star-formation features in the host
galaxies (e.g. Djorgovski et al. \cite{Dj98}). Emission and absorption
features observed in X-ray spectra (e.g. Piro et al. \cite{Pi00}) are
also indicative of large amounts of metals in the GRB surroundings,
which can be provided by SN explosions. Finally, the afterglow spectra
of GRB\,021004 showed absorption systems at different velocities,
suggesting the presence of rapidly-moving pre-ejected material
surrounding the explosion site (e.g. M\o{}ller et al. \cite{Mo03}).

\begin{figure}
\epsfig{file=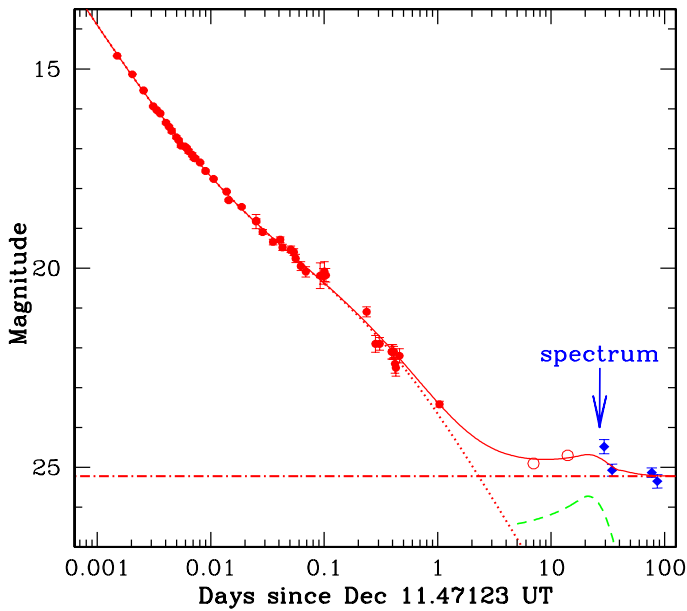, width=9.cm, height=8.0cm, angle=0}
\vskip 0.0truecm
\caption{Light curve of the afterglow of GRB\,021211. Filled circles
  represent data from published or submitted works (Fox et
  al. \cite{Fo03a}; Li et al. \cite{Li03}; Pandey et al. \cite{Pa03}),
  open circles are converted from HST measurements (Fruchter et
  al. \cite{Fr02}), while filled diamonds indicate our data; the arrow
  shows the epoch of our spectroscopic measurement. The dotted and
  dot-dashed lines represent the afterglow (see text) and host
  contribution respectively. The dashed line shows the light curve of
  SN\,1994I reported at $z = 1.006$ and dereddened with $A_V = 2$ (from
  Lee et al. \cite{Le95}). The solid line shows the sum of the three
  contributions.\label{fg:lc}}
\end{figure}

\section{GRB\,021211.}\label{sec:GRB}

GRB\,021211 was detected on 2002~Dec~12 by the \mbox{HETE--2} satellite
at 11:18:34~UT (Crew et al. \cite{Cr02}). The $\gamma$-ray fluence was
$(0.96 \pm 0.29) \times 10^{-6}$~erg~cm$^{-2}$ in the ($7 - 30)$~keV
band and $(1.98 \pm 0.15) \times 10^{-6}$~erg~cm$^{-2}$ in the $(30 -
400)$~keV band (Crew et al. \cite{Cr03}). The GRB was therefore
classified as X-ray rich. Given the redshift $z = 1.006$ of this event
(Vreeswijk et al. \cite{Vr02a}), the inferred (isotropic) energy was
$E_\mathrm{iso} = (6 \pm 0.5) \times 10^{51}$~erg, at the low end of the
energy distribution of GRBs (Frail et al. \cite{Fr01}).  Owing to the
rapid distribution of coordinates, an optical afterglow was rapidly
discovered by Fox \& Price (\cite{FP02}), just 20~minutes after the GRB
onset, as a pointlike source with magnitude $R = 18.29 \pm 0.02$ at
coordinates $\alpha = 08^\mathrm{h} 08^\mathrm{m} 59\fs9$, $\delta =
+06\degr 43\arcmin 37\farcs5$ (J2000; Fox et
al. \cite{Fo03a}). Moreover, the automatic telescopes RAPTOR (Wozniak et
al. \cite{Wo02}), KAIT (Li et al. \cite{Li03}), and Super--LOTIS (Park
et al. \cite{Pa02}), imaged the error box a few minutes after the
GRB. These observations allowed to monitor the early lightcurve of the
afterglow, which could be described by a broken powerlaw with a
flattening at $t \approx 10$~min (Li et al. \cite{Li03}).  The striking
feature of the optical afterglow of GRB\,021211 was however its extreme
faintness. Compared to other events at similar epochs and redshifts,
this afterglow was dimmer by $\sim 3$~magnitudes in the $R$-band (see
e.g. Fox et al. \cite{Fo03a}).  This could in principle be due to heavy
extinction within the host. However, Fox et al. (\cite{Fo03a}) report a
broad-band color $B-K = 3.9$ (on Dec~11.6~UT), well within the range of
GRB afterglows (\v{S}imon et al. \cite{Si01}). This indicates that
GRB\,021211 suffered negligible extinction. Moreover, this burst was
also underluminous in all observed wavebands: only upper limits were
reported in the radio (Fox et al. \cite{Fo03a}; Rol \& Strom
\cite{RS02}), submillimeter (Fox et al. \cite{Fo03a}; Hoge et
al. \cite{Ho02}), and TeV (McEnery
\cite{ME02}) regions. Unfortunately, no follow-up X-ray observations
were performed.

The intrinsic faintness of the optical afterglow made this event a good
candidate for showing a prominent late-time bump.

\begin{figure*}
\epsfig{file=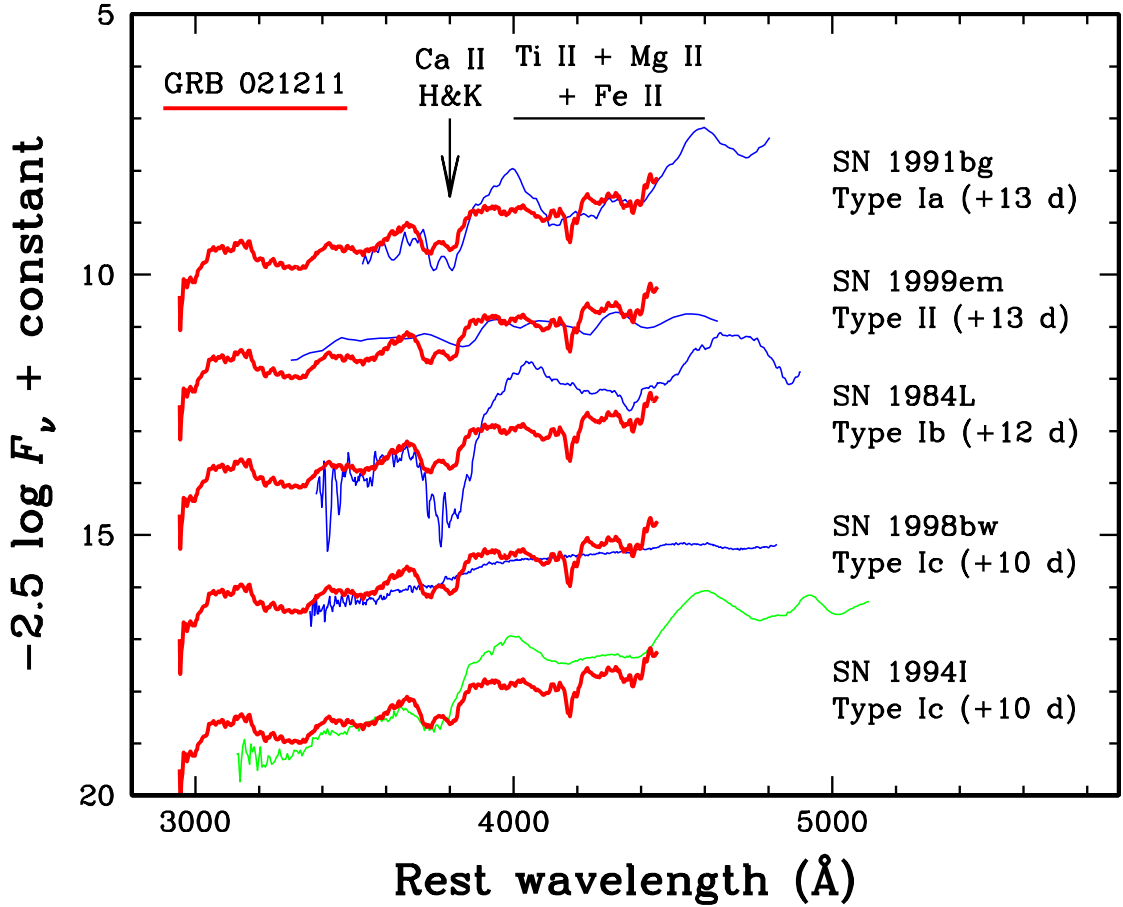, height=12cm}
\vskip 0.0truecm
\caption{Rest-frame spectrum of the afterglow of GRB\,021211 on
  2003~Jan~8.27~UT, or 27~days after the GRB (thick lines), compared
  with that of several SNe (thin lines).\label{fg:spec}}
\end{figure*}

\section{Photometry.}\label{sec:data}

\begin{table}
\caption{Summary of our photometric observations. Magnitudes are
referred to the complex afterglow + host. Errors are 1-$\sigma$. $^*$:
average of two measurements obtained on Jan~9.3 and
10.2.\label{tb:photo}}
\centering\begin{tabular}{llll}\hline
  Start UT         & Exp time & Seeing      & $R$ magnitude   \\ \hline
  2003 Jan 9.7$^*$ & 63 min   & 1.1\arcsec  & $24.48\pm0.18$  \\
  2003 Jan 15.33   & 15 min   & 0.8\arcsec  & $25.07\pm0.15$  \\
  2003 Feb 28.02   & 75 min   & 0.8\arcsec  & $25.13\pm0.12$  \\
  2003 Mar 9.01    & 50 min   & 1.25\arcsec & $25.35\pm0.17$  \\ \hline
\end{tabular}
\end{table}

Late-time observations were secured at the ESO VLT--UT4 (Yepun) equipped
with the FORS\,2 instrument, in the $R$ band, during the period January
-- March 2003 (see Tab.~\ref{tb:photo}).

The conversion to absolute flux was obtained by using both standard
calibration and a secondary sequence calibrated, on Feb~28, with a Landolt
standard field (SA98), to account for the observations
obtained under non-photometric conditions. Aperture photometry was
executed with the packages \texttt{apphot} and \texttt{photcal} within
IRAF, by choosing aperture radii from 0.5\arcsec{} to 3\arcsec, then
correcting to infinity with the values found for the standards and
using the \texttt{DAOGROW} algorithm. The measurements on the various
apertures for the secondary calibrators were found very stable, and we
decided to adopt for the target the innermost aperture magnitude,
corrected for the aperture.

Our results are listed in Tab.~\ref{tb:photo} and supersede our
preliminary report (Testa et al. \cite{Te03}). They have been
complemented with a compilation of observations collected from
literature and plotted in Fig.~\ref{fg:lc}. A rebrightening is clearly
seen, starting $\sim 15$~days after the burst (Fruchter et
al. \cite{Fr02}) and reaching the maximum, $R \sim 24.5$, during the
first week of January. The contribution of the host galaxy, estimated
from our late-epoch images, is $R = 25.22 \pm 0.10$. Therefore, the
intrinsic magnitude of the bump was $R = 25.24 \pm 0.38$.

The afterglow contribution is more uncertain. The early-time light curve
presents several fluctuations, and cannot be easily extrapolated to
later epochs. In particular, fitting all data up to 1~day after the
GRB\footnote{We added in quadrature 0.03~mag to the errors of all points
in order to account for the use of different telescopes and
calibrations.} with a (convex) broken powerlaw, as used by Li et
al. (\cite{Li03}), yields an unacceptable $\chi^2/\dof = 115/45$.  We
suggest two possible alternatives. First, the shape of the lightcurve is
consistent with the presence of two or three rebrightenings underlying a
powerlaw component, similarly to what observed in GRB\,021004 (Lazzati
et al. \cite{La02}; Nakar et al. \cite{Na03}). Alternatively, a second
break in the light curve could be present about 10~hours after the
GRB. This may be the consequence of the passage of the synchrotron
cooling frequency across the optical band (the jet break is expected to
occur much later for such a low energy event, according to the
correlation of Frail et al. \cite{Fr01}). A fit with a double broken
powerlaw (i.e. with {\em two} breaks, see Fig.~1) yields $\chi^2/\dof =
55/43$; the improvement is highly significant ($F$-test chance
probability less than $10^{-6}$). We note that, whichever extrapolation
is used, the contribution of the afterglow to the flux measured at the
epoch at which our spectrum (the arrow in Fig.~\ref{fg:lc}) was
obtained, is negligible (less then $5\%$ in the most conservative
case). This fact strongly supports that the bump is powered by a
different component other than the afterglow.

\section{Spectral Analysis.}\label{sec:spectrum}

We obtained a spectrum of the afterglow + host with FORS\,2, on
Jan~8.27~UT (27 days after the GRB), during the rebrightening phase
shown in Fig.~\ref{fg:lc}. The original spectrum covered the range of
wavelengths $(6000 - 11000)$~\AA, although only the interval $(6000 -
9000)$~\AA{} afforded an acceptable S/N ($\gtrsim 3$).  The resolution
was about 19~\AA, and the integration time was $4 \times 1$~h. The slit
was rotated in such a way to include also the nearby galaxy reported by
Caldwell et al. (\cite{Ca02}); this object was well detected and is
clearly separated from our target (the seeing in the 4 exposures was
$0\farcs6 - 1\farcs4$). The extraction of the spectrum was performed
within the MIDAS and IRAF environments, independently by three of us
(S.B., M.H., and M.D.V.). We clearly detected the emission line at
7472.9~\AA{} already found by Vreeswijk et al.
(\cite{Vr02a}). Interpreting this as [O\,II] 3727~\AA{} in the rest
frame of the host galaxy, this corresponds to a redshift $z =
1.006$. Following this idea, we could also detect, albeit at lower
significance, emission lines at 9720~\AA, 9955~\AA, and 10025~\AA, which
we intepret as H$\beta$ and [O\,III] 4959~\AA{} and 5007~\AA. Our
spectrum, shown in the rest-frame of the GRB (thick lines in
Fig.~\ref{fg:spec} and Fig.~\ref{fg:spec_large}) was smoothed with a
boxcar filter (55~\AA{} width) and cleaned from the emission line
[O\,II]. The spectrum of the afterglow is characterized by broad
low-amplitude undulations blueward and redward of a broad absorption,
the minimum of which is measured at $\sim 3770$~\AA{} (in the rest frame
of the GRB), whereas its blue wing extends up to $\sim 3650$~\AA.  We
then compared our spectrum with those of SNe arranged in different
spectroscopic types and obtained at different epochs
(Fig.~\ref{fg:spec}, thin lines). The comparison includes the type-Ia
SN\,1991bg (Filippenko et al. \cite{Fi92}; Turatto et al. \cite{Tu96}),
the type-II SN\,1999em (Hamuy et al. \cite{Ha01}), the type-Ib SN\,1984L
(Harkness et al. \cite{Ha87}), the peculiar type-Ic SN\,1998bw (Galama
et al. \cite{Ga98}; Patat et al. \cite{Pa01}), and the type-Ic SN\,1994I
(Filippenko et al. \cite{Fi95}). Both SN\,1999em and SN\,1998bw provide
a poor match to the afterglow spectrum. Some similarity can be found
with the type-Ia SN\,1991bg and the type-Ib SN\,1984L. A more convincing
resemblance is found with the spectrum of the type-Ic SN\,1994I,
obtained 9~days after its $B$-band maximum (Filippenko et
al. \cite{Fi95}). The comparison with SN\,1994I (and to some extent also
with SN\,1991bg and SN\,1984L) strongly supports the identification of
the broad absorption with Ca\,II H+K; the blueshifts corresponding to
the minimum of the absorption and to the edge of the blue wing imply
velocities of $v\sim 14\,400$~km/s and $v\sim 23\,000$~km/s
respectively. In principle the Ca\,II from the host galaxy could
contaminate our spectrum. However, since the typical FWHM of absorption
(and emission) lines of galaxies is of the order of $10 - 15$~\AA{}
(corresponding to less than $1\,000$~km/s), and the FWHM of the observed
absorption is about 150~\AA, one concludes that the Ca\,II of the host
galaxy might affect the observed feature only marginally, if at all.

\section{Measurements of the Time Delay between the SN and the GRB.}
\label{sec:consistency}

\begin{figure}
\epsfig{file=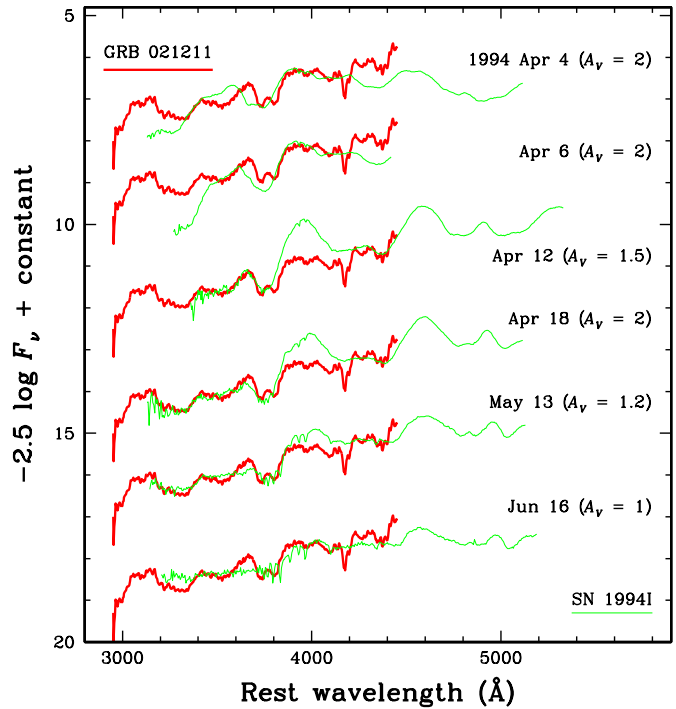, width=\columnwidth}
\vskip 0.0truecm
\caption{Comparison of the spectrum of the GRB bump to the ones of
  SN\,1994I taken at different times. Each plot has a different $A_V$,
  in order to better match our spectrum (we adopted the extinction law
  by Cardelli et al. \cite{CCM89}). For reference, SN\,1994I reached its
  $B$-band maximum on 1994~Apr~9.\label{fg:time}}
\end{figure}

Using SN\,1994I as a template, our photometric and spectroscopic data
allow us to estimate the time at which the SN exploded, and to compare
it with the GRB onset time.

{\bf Spectroscopy.} In order to further study the spectrum of the SN
associated with the bump, we have used as a reference (see
Fig.~\ref{fg:time}) the spectral evolution of SN\,1994I (Filippenko et
al. \cite{Fi95}). To find the best match, different reddenings in the
range $A_V = 1.4 \pm 0.5$ (Richmond et al. \cite{Ri96}) were applied to
each spectrum. The epochs before maximum light (April~9) seem to be
excluded by the morphology of Ca feature, which is broader and located
at shorter wavelengths than in the GRB (the correctness of the
wavelenght axis was checked by measuring the position of the NaD
interstellar absorption in each spectrum). Spectra later than Jun~1 show
a very weak Ca feature and exhibit a poor match with the afterglow
spectrum in the ($3200 - 3700$)~\AA{} region. Therefore the match with
the spectra is acceptable over the range $\sim\mbox{}$Apr~9 to
$\sim\mbox{}$May~31 ($\sim\mbox{}$0 to 40 days after SN maximum light).

Since our spectrum was obtained on Jan~8, we estimate that the maximum
of the SN should have occurred between 2002~Oct~2 and 2003~Jan~8 (since
the source was at $z = 1$). The exact epoch when the SN exploded depends
crucially on the rise time (the time interval from the epoch of the
explosion up to maximum light) of type-Ic SNe.  The best documented
cases are SN\,1998bw and SN\,1999ex. The latter (Stritzinger et
al. \cite{St02}) reached $B$-band maximum $\sim 18$~days after the
explosion, the former after $\sim 16$ days (Galama et
al. \cite{Ga98}). SN\,1994I had a faster rise, reaching its maximum (in
$B$) only 12 days after the explosion (Iwamoto et
al. \cite{Iw94}). Adopting the latter value, we conclude that the SN
exploded between $\sim\mbox{}$Sept~11 and Dec~15, the later epochs being
favored.

{\bf Photometry.} In Fig.~\ref{fg:lc} we have superimposed to the light
curve of the afterglow decay the light curve of SN\,1994I (Lee et
al. \cite{Le95}), reported\footnote{Assuming $\Omega_\mathrm{m} = 0.3$,
$\Omega_\Lambda = 0.7$, $H_0 = 71$~km~s$^{-1}$~Mpc$^{-1}$, the distance
modulus at $z=1.006$ is $\mu = 44.07$~mag.} at $z = 1.006$ and
dereddened by $A_V = 2$~mag (Richmond et al. \cite{Ri96}). The
$K$-correction has been computed from $U$-band data, considering that,
at $z = 1.006$, the $U$-band roughly corresponds to the observed
$R$-band. The plot (solid line) shows that the luminosity at maximum of
SN\,1994I ($M_U = -18.9 \pm 0.3$ assuming $A_V = 2$; see Tab.~10 of
Richmond et al. \cite{Ri96}) agrees very well with that of the bump
($M_U = -18.8 \pm 0.4$). In the figure, a null time delay between the
GRB and the SN explosion was used. Letting this delay free to vary did
not significantly improve the fit ($F$-test chance probability of
$36\%$); the best fit time delay is $t_\mathrm{GRB}-t_\mathrm{SN} =
(-1.5 \pm 3)$~comoving days.

Evidently the photometric observations provide a tighter constraint on
the SN-GRB delay. However the uncertainties above are only statistical,
while systematics are more difficult to evaluate, especially in
consideration of the paucity of observation of type-Ic SNe in their
rising phase. Yet, the combination of our photometric and spectroscopic
data provide evidence that the SN and the GRB explosions occurred
within days from one another, at the most.

\section{Conclusions.}\label{sec:concl}

The detection of a broad (FWHM $\sim 150$~\AA) absorption feature, in
the spectrum of the `bump', which we have identified with the Ca\,II H+K
doublet (blueshifted by $\sim 15\,000$ km/s) suggests that the
rebrightening of the GRB\,021211 afterglow was powered by a SN. Assuming
for this SN a spectroscopic and photometric behavior similar to that of
SN\,1994I, our data indicate that the SN and GRB explosion may have
occurred almost simultaneuosly, or at most separated by a few days.  The
temporal coincidence between the SN and GRB\,021211 holds for `short'
rise times, of the order of $10 - 12$~days (as observed for SN 1994I;
Iwamoto et al. \cite{Iw94}).  On the other hand, if a longer rise time
($16 - 20$~days, such as that observed in SN\,1998bw or SN\,1999ex) were
used, the conclusion would be that the SN went off several days before
GRB\,021211.

It is interesting to note that SN\,1994I, the spectrum of which provides
the best match to that observed in GRB\,021211, is a typical type-Ic
event rather than an exceptional 1998bw-like object, as the one proposed
for association with GRB\,980425 and GRB\,030329 (Galama et
al. \cite{Ga98}; Stanek et al. \cite{St03}). If the SN associated with
GRB\,021211 indeed shared the properties of SN\,1994I, this would open
the interesting possibility that GRBs may be associated with standard
type-Ic SNe, and not only with the more powerful events known as
`hypernovae'. This fact may have interesting consequences on the rate of
GRB events. One caveat is that the recently studied SN\,2002ap (Mazzali
et al. \cite{Ma02}) shared some of the properties of hypernovae (e.g. a
high expansion velocity), but was not significantly brighter than
standard type-Ic SNe.  Even if its pre-maximum spectra showed
significantly broader lines than our case, this difference vanished
after maximum, such that it may not be easy to distinguish between the
two types of SNe. However, SN\,2002ap had a broader light curve, and it
was too faint in the $U$-band (Yoshii et al. \cite{Yo03}). It remains
however not firmly estabilished whether GRB\,021211 was associated with
a standard type-Ic SN or with a `low energy' hypernova similar to
SN\,2002ap. We last note that even if GRBs are indeed mainly associated
with standard type-Ic SNe, the discovery of overluminous type-Ic events
(like SN\,1998bw) associated with GRBs is observationally favored, since
the SN component can emerge and be observed at early times, when the
transient is more frequently monitored.

\begin{acknowledgements}
We thank Alexei Filippenko and Nando Patat for giving us the spectra of
SN\,1994I and SN\,1998bw respectively, and the ESO directorate for
accepting our DDT proposal. Mario Vietri, Peter H\"oflich and Lucia
Ballo are acknowledged for useful discussion. We also thank the
anonymous referee for her/his comments, which have improved the
presentation of the data. We also appreciate the work of the observing
staff at Paranal. We thank the Italian Collaboration for Optical Bursts
(CIBO) for showing us preliminary photometric data.
\end{acknowledgements}

\begin{figure*}
\includegraphics[width=\textwidth]{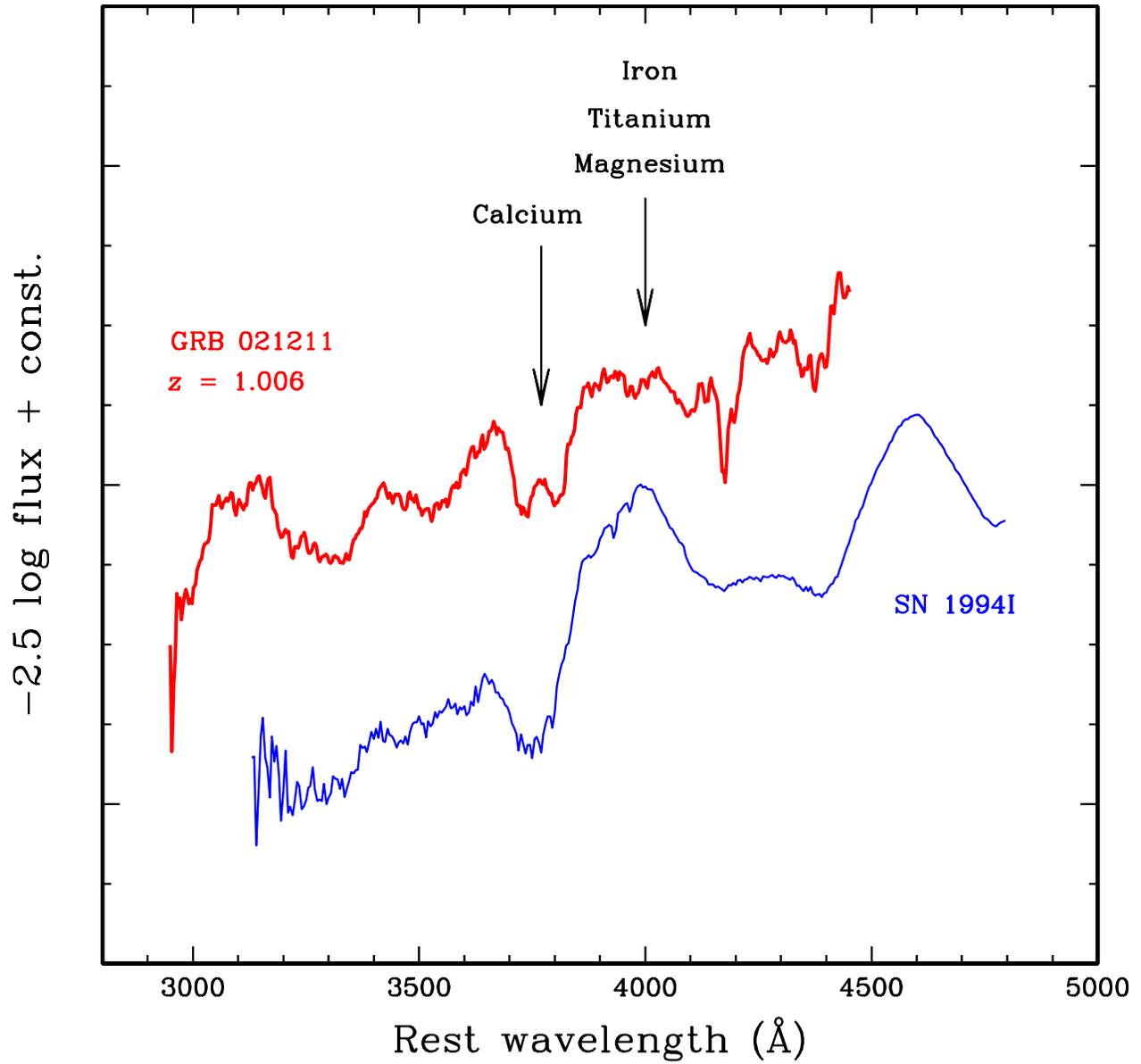}
\caption{The thick line showe the rest-frame spectrum of the afterglow
  of GRB\,021211 on 2003~Jan~8.27~UT, compared with that of SN 1994I
  (thin line).\label{fg:spec_large}}
\end{figure*}

\end{document}